\documentclass[conference,a4paper]{IEEEtran}

\usepackage{epsfig,makeidx,color}
\usepackage{graphicx}
\usepackage{amsbsy}
\usepackage{amsmath}
\usepackage{amssymb}
\usepackage{euscript,verbatim}
\newtheorem{exam}{Example}

\newcommand{\Z}{\mathbf{Z}} % integer set
\newcommand{\R}{\mathbf{R}} % real set
\newcommand{\Q}{\mathbf{Q}} % rational set
\newcommand{\diag}{{\hbox{diag}}}
\renewcommand{\det}{{\hbox{\rm det}}}

\begin{document}

\title{Rotating Non-Uniform and High-Dimensional Constellations Using Geodesic Flow on Lie Groups}

\author{
  \IEEEauthorblockN{David A. Karpuk}
  \IEEEauthorblockA{Dept. of Mathematics and Systems Analysis\\
    Aalto University\\
    P.O.\ Box 11100\\
    FI-00076 Aalto, Finland\\
    email: david.karpuk@aalto.fi} 

\and 

  \IEEEauthorblockN{Camilla Hollanti}
  \IEEEauthorblockA{Dept. of Mathematics and Systems Analysis\\
    Aalto University\\
    P.O.\ Box 11100\\
    FI-00076 Aalto, Finland\\
    email: camilla.hollanti@aalto.fi} 

}

\maketitle

\begin{abstract}
We use a numerical algorithm on the Lie group of rotation matrices to obtain rotated constellations for Rayleigh fading channels.  Our approach minimizes the union bound for the pairwise error probability to produce rotations optimized for a given signal-to-noise ratio. This approach circumvents explicit parametrization of rotation matrices, which has previously prevented robust numerical methods from being applied to constellation rotation. Our algorithm is applicable to arbitrary finite constellations in arbitrary dimensions, and one can thus apply our method to non-uniform constellations, which are of interest for practical concerns due to their ability to increase BICM capacity.  We show how our rotations can improve the codeword error performance of non-uniform constellations, and we also apply our method to reproduce and improve rotations given by ideal lattices in cyclotomic fields.

\end{abstract}

\begin{IEEEkeywords}
Rayleigh fading channel, rotated constellations, NUQAM constellations, ideal lattices
\end{IEEEkeywords}

\IEEEpeerreviewmaketitle

\section{Introduction}

Transmission over a wireless channel typically involves sending a signal which is subsequently affected by fading and noise. Ensuring that such transmission is done reliably requires an energy-efficient constellation resistant to both of these effects. It is known that rotating constellations is an effective way of combating the effect of fading, while maintaining the energy of each codeword.

We consider transmission over a Rayleigh fading channel.  As in \cite{OV}, by using a bit interleaver, assuming perfect channel state information at the receiver, and separating real and imaginary parts, we can reduce the model to
\begin{equation}\label{model}
y = Hx + z
\end{equation}
where 
\begin{itemize}
\item[$\bullet$] $x\in\R^n$ is the codeword intended for transmission,
\item[$\bullet$] $H=\diag(\alpha_i)$ is a real diagonal $n\times n$ matrix with $\alpha_i$ a Rayleigh distributed random variable with $\mathbf{E}(\alpha_i^2) = 1$,
\item[$\bullet$] $z=(z_i) \in\R^n$ a noise vector with $z_i$ a real zero-mean Gaussian random variable with variance $\sigma_e^2$, and 
\item[$\bullet$] $y\in \R^n$ the received vector.
\end{itemize}

Codebooks for Rayleigh fading channels traditionally consist of uniformly spaced $Q$-QAM symbols, i.e.\ a finite constellation $\mathcal{C}\subset \R^2$ of size $Q=2^q$.  For example, let us consider the $4$-QAM constellation $\{(\pm1,\pm1)\}$. A deep fade (some $\alpha_i\approx 0$) essentially erases one of the coordinates from our codewords, thereby making it potentially impossible for the receiver to distinguish between, for example, $(1, 1)$ and $(1,-1)$, in the case where the second coordinate is erased.  Rotating a constellation can combat a deep fade by providing diversity among the coordinates of the codewords.  For sufficiently large SNR, the \emph{minimum product distance}
\begin{equation}
d_{p,\min} = \min_{x\neq y}\prod_{x_i\neq y_i}|x_i-y_i|
\end{equation}
of a constellation provides useful design criteria, and one seeks a rotation which maximizes the minimum product distance over all pairs of constellation points.  

However, the minimum product distance is only an asymptotic design criteria.  In contrast, we present a numerical scheme which takes as input a finite constellation $\mathcal{C}\subset\R^n$ \emph{and a specific} SNR, and outputs a rotation matrix optimal in terms of minimizing the pairwise error probability \emph{for that level of} SNR.  We do this by performing numerical analysis on the Lie group $SO(n)$ of all $n$-dimensional rotation matrices.  Our approach is relevant for low- and mid-range SNR levels, and is therefore applicable to many of the practical scenarios considered by, for example, the Digital Video Broadcasting (DVB) consortium.

The DVB consortium has established rotated QAM constellations as a part of the DVB-T2 industry standard \cite{DVBT2}, and NUQAM (non-uniform QAM) constellations have been considered in DVB-NGH (next-generation handheld) implementation \cite{DVBNGH}.  A natural question is therefore whether rotations can increase the performance of NUQAM constellations.  Such constellations may lack any algebraic structure, and thus do not support algebraic or number theoretic techniques when attempting to obtain effective rotations. 

%The goal of this paper is to provide a numerical scheme for obtaining optimal rotations for any finite constellation $\mathcal{C}\subset\R^n$ and any $n$.  Numerical methods for rotating constellations were abandoned early on due to the difficulty of parametrizing high-dimensional rotation matrices [6].  However, our method avoids explicitly parametrizing rotation matrices, resulting in an algorithm more easily applicable to higher-dimensional constellations.  

In summary, the contributions of this paper lie within
\begin{itemize} 
\item[$\bullet$] applying geodesic flow on the Lie group of rotation matrices to the problem of rotating constellations for Rayleigh fading channels,
\item[$\bullet$] using our method to improve the codeword error rate performance of non-uniform QAM (NUQAM) constellations,
\item[$\bullet$] demonstrating how to slightly improve the codeword error rate of the DVB-NGH recommended four-dimensional rotation matrix, for $4$D QAM constellations,
\item[$\bullet$] and reproducing and improving upon some algebraically rotated lattices coming from cyclotomic number fields.
\end{itemize}

\section{Related Work}

The idea of rotating two-dimensional constellations to obtain an increase in diversity was first presented in \cite{boulle}.  Since then, analytical expressions for optimal rotations of two-dimensional constellations were presented in \cite{wangwang,kiyani}, and a numerical scheme for optimizing over low-dimensional rotations was presented in \cite{jelicec}.  Numerous algebraic and number theoretic techniques also exist to construct fully-diverse lattices with good minimum product distance \cite{OV}, and have been shown to outperform randomly rotated QAM constellations in terms of bit error rate \cite{fabregas}.  Non-uniform constellations which approximate samplings from a Gaussian distribution have been studied in \cite{verduwu,betts} due to their ability to improve capacity.

Previous methods have assumed sufficiently large SNR and determined rotations which optimize with respect to the minimum product distance.  Our work differs from the vast majority of the previous literature by  optimizing with respect to a more accurate estimate of the pairwise error probability.  Our method also locates rotations which are optimal for a given SNR, rather than sufficiently large SNR.  Additionally, our method does not rely on any algebraic structure, and thus can be applied to rotate non-uniform constellations as well.  Lastly, previous numerical methods have all been based on an explicit parametrization of rotation matrices of a given dimension, while working with the Lie groups and Lie algebras avoids this problem.

\section{Constellation Rotations for Rayleigh Fading Channels}\label{rot}

All constellations $\mathcal{C}\subset\R^n$ we consider will be normalized to have unit average energy, that is, we scale $\mathcal{C}$ by a constant factor so that
\begin{equation}
E_s = \mathbf{E}(||x||^2) = 1
\end{equation}
where the expectation is taken over all $x\in\mathcal{C}$.  We then define the signal-to-noise ratio by
\begin{equation}
\text{SNR} = E_s/N_0 = 1/N_0
\end{equation}
though all simulation results will use the quantity $10\log_{10}(\text{SNR})$ to measure signal-to-noise ratio in decibels.

Let $\mathcal{C}\subset \R^n$ be a finite constellation.  The resulting pairwise error probability of the channel model (\ref{model}) is often estimated by the quantity
\begin{equation}\label{prob}
P_e \leq C \sum_{x,y\in\mathcal{C}}\prod_{i=1}^n \frac{1}{1+\frac{|x_i-y_i|^2}{8N_0}},
\end{equation}
where $C$ is a constant independent of $\mathcal{C}$; see \cite{divsalar,boutrousgood}.  Let $Q$ be a rotation matrix, that is, $QQ^t=I_n$ and $\det(Q)=1$.  We will denote the constellation obtained by applying $Q$ to each point of $\mathcal{C}$ by $\mathcal{C}_Q$.  To find a rotation which minimizes the pairwise error probability, we will find a local minimum of the objective function
\begin{equation}\label{obj}
\boxed{f(Q) = \sum_{x,y\in\mathcal{C}_Q}\prod_{i=1}^n \frac{1}{1+\frac{|x_i-y_i|^2}{8N_0}}}
\end{equation}
in the space of all rotation matrices.

One often bounds (\ref{obj}) by assuming $N_0$ is sufficiently small, which results in studying the minimum product distance of the constellation, especially when $\mathcal{C}$ is a carved from a lattice.  However, the expression (\ref{obj}) exhibits smoother behavior with respect to rotation than the minimum product distance, and is thus more amenable to numerical optimization, and additionally is a more accurate estimate.

\section{Lie Groups and Lie Algebras}\label{graddescent}

Finding an optimal rotation will require performing numerical analysis on the Lie group $SO(n)$, the space of all $n$-dimensional rotation matrices.  Because $SO(n)$ is not a Euclidean vector space, traditional steepest descent techniques cannot be used.  In this section we provide the necessary basics of the theory of Lie groups needed to describe the numerical algorithm we use.  As a general reference for the theory of Lie groups we recommend \cite{hallie}.

A \emph{Lie group} $G$ is both a manifold and a group, such that the group operations of multiplication and inversion on $G$ are continuous with respect to the manifold structure.  For example, the group $GL(n)$ of $n\times n$ invertible real matrices is a Lie group, in which the group operations are the normal matrix operations.  Such operations are clearly continuous, since they are polynomials in the matrix entries.

We are interested in the Lie group $SO(n)$ of $n\times n$ rotation matrices, defined by
\begin{equation}
SO(n) = \{Q\in GL(n)\ |\ QQ^t = I_n,\ \det(Q) = 1\}.
\end{equation}
One can show that $SO(2)$ is the set of matrices of the form
\begin{equation}\label{2drot}
\left(\begin{array}{rr}
\cos(\theta) & -\sin(\theta)\\
\sin(\theta) & \cos(\theta)
\end{array}\right)
\end{equation}
for $\theta \in \R$.  Such a matrix corresponds to rotating $\R^2$ counterclockwise by $\theta$ radians, when vectors are multiplied on the right.  Such an explicit parametrization becomes more difficult as $n$ increases.  The dimension of $SO(n)$ (the minimum number of parameters required to describe such a matrix) is $(n^2-n)/2$, whereas the dimension of $GL(n)$ is $n^2$.

The \emph{Lie algebra} $\frak{g}$ of a Lie group $G$ is the tangent space at the identity element $e\in G$, and thus is a real vector space of dimension equal to that of $G$.  The Lie algebras of $GL(n)$ and $SO(n)$ have convenient explicit descriptions:
\begin{eqnarray}
\frak{gl}(n) &=& M(n)\\
\frak{so}(n) &=& \{A\in M(n)\ |\ -A = A^t\}
\end{eqnarray}
where $M(n)$ is the set of all $n\times n$ real matrices.  One can pass from the Lie algebra to the Lie group using the exponential map
\begin{equation}
\exp:\frak{g}\rightarrow G,
\end{equation}
defined by the familiar power series
\begin{equation}
\exp(A) = I_n + A + \frac{A^2}{2!} + \frac{A^3}{3!}+\cdots
\end{equation}
for Lie groups $G$ which are subgroups of $GL(n)$. The exponential of a matrix can be computed numerically in MATLAB.

\section{Geodesic Flow on $SO(n)$}

Let $f:M(n)\rightarrow \R$ be an objective function which we wish to minimize on the subset $SO(n)\subset M(n)$.  On the whole of $M(n)$ one can use a standard gradient descent procedure, however there is then no guarantee that the final output $Q_N$ is a rotation matrix.  We thus require a gradient descent procedure which starts with a rotation matrix $Q_0$ and maintains the constraints $QQ^t,\ \det(Q) = 1$ while moving about the search space.  To do this, we use the method of \emph{geodesic flow}, introduced in this numerical form by Nishimori in \cite{nishimori}, and summarized nicely by Plumbley in \cite{plumbley}.  The geodesic flow algorithm approximates a path in $SO(n)$ between the matrices $Q_0$ and $Q_N$, by taking successive infinitesimal steps in a direction determined by the gradient on the Lie algebra.

The method of geodesic flow starts with a rotation matrix $Q_0$ and a step size $h$, and recursively updates via the formula
\begin{equation}
Q_{k+1} = \exp(-h\frak{r}_k)Q_k
\end{equation}
where
\begin{equation}
\frak{r}_k = (\nabla f|_{Q_k})Q_k^t - Q_k(\nabla f|_{Q_k})^t.
\end{equation}
The matrix $\frak{r}_k$ is skew-symmetric and should be thought of as an infinitesimal gradient matrix living on the Lie algebra $\frak{so}(n)$.  Thus $\exp(-h\frak{r}_k)$ is an element of $SO(n)$, which restricts this gradient search algorithm to the space of rotation matrices as desired.  For an explanation of why this algorithm locates a local minimum of $f$, see \cite{plumbley}.

Of course, a nice analytic expression for $\nabla f$ may not exist; therefore we use a central difference approximation.  Let us define the matrices
\begin{equation}
H_{ij} = \left\{ \begin{array}{cl}
h & \text{in the $ij$th entry}\\
0 & \text{elsewhere}
\end{array}\right.
\end{equation}
for $1\leq i,j\leq n$.  For a fixed $A\in M(n)$, we approximate the $ij$th entry of the gradient of $f$ at $A$ by
\begin{equation}
(\nabla f|_A)_{ij} \approx \frac{f(A+H_{ij})-f(A-H_{ij})}{2h}.
\end{equation}
One can of course use higher-order difference approximations to obtain a more accurate numerical scheme.

\section{Examples and Simulation Results}

We now apply the geodesic flow method to minimize the function $f$ of (\ref{obj}), for certain non-uniform and high-dimensional constellations.  For simplicity, we fix for all of our simulations a step size of $h=.01$ and run our optimization routine over $N=10000$ trials.  While line search algorithms and a carefully chosen stopping condition could improve our scheme, our emphasis is more on proof-of-concept than technical numerical methods.

\subsection{$2$D $16$-NUQAM Constellations}

In this section, we study how our method can improve the performance of a two-dimensional NUQAM constellation $\mathcal{C}\subset\R^2$.  For the sake of simplicity we will restrict our attention to the case of 16-NUQAM constellations, whose coordinates are taken from the set $\{\pm1,\pm\gamma\}$ where $\gamma > 1$.

The parameter $\gamma$ is selected to maximize the BICM capacity for a fixed SNR \cite{douillard}, and as $\text{SNR}\rightarrow\infty$, one can show that the optimal $\gamma\rightarrow3$, so that optimal 16-NUQAM converges to the traditional 16-QAM.  Here we choose $\gamma = 3.15$, and study how it and its corresponding optimal rotations perform at higher SNR levels. 

Figure $1$ below shows that our rotations outperform the corresponding unrotated constellation.  We also plot the corresponding error curve for a rotation by $16.8^\circ$, the rotation recommended by DVB for $16$-QAM.  While the choice of angle was made with uniform constellations in mind, it nevertheless provides a useful control for our experiment.

\begin{figure}[h!]
\hspace{-.02\textwidth}
\includegraphics[width = .5\textwidth]{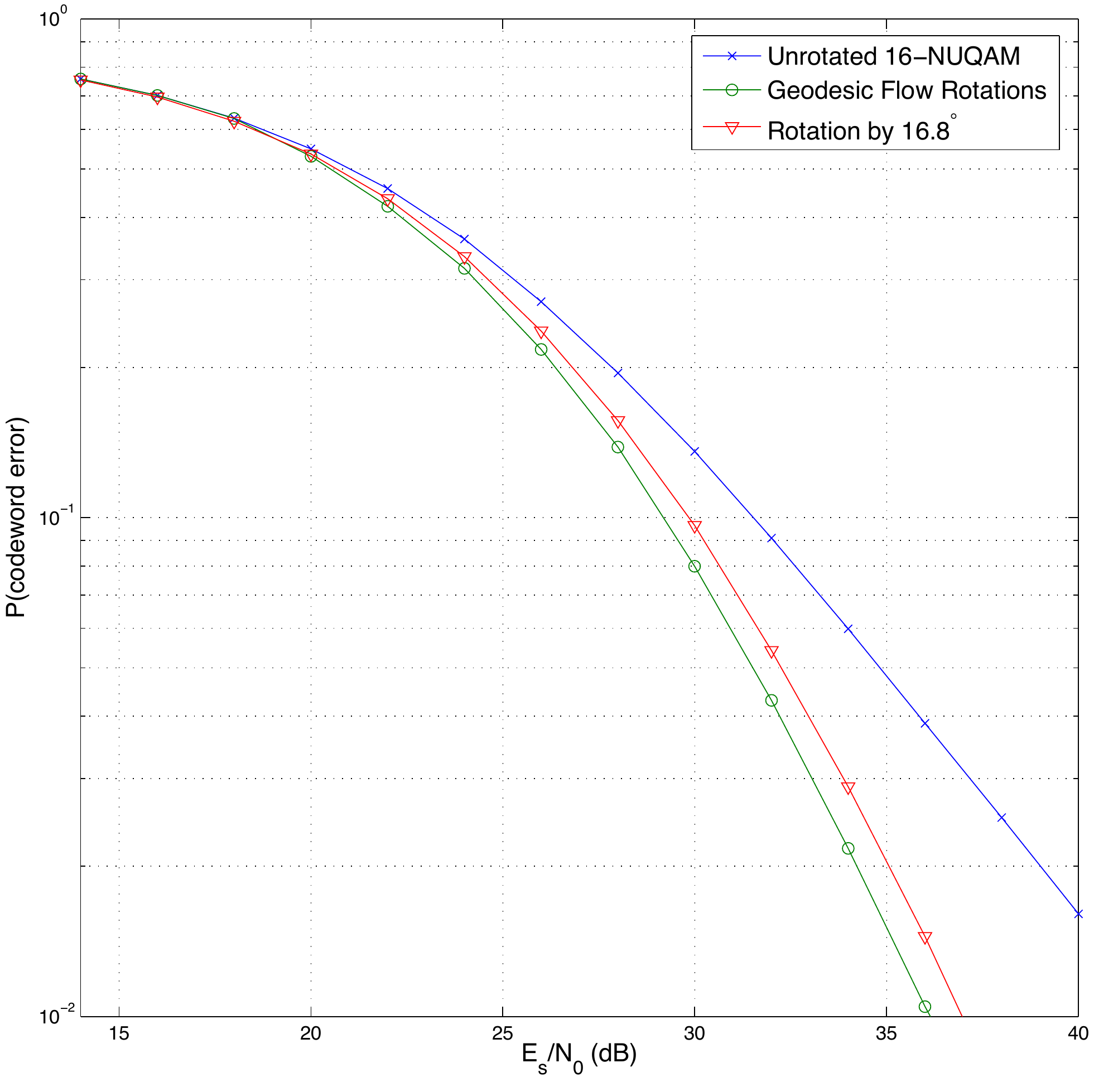}
\vspace{-.04\textwidth}\caption{Codeword Error Rates of 16-NUQAM Constellations}
\end{figure}

\subsection{$4$D $4$-QAM Constellations}

We will now apply our method to rotate the constellation
\begin{equation}
\mathcal{C} = \{(\pm1,\pm1,\pm1,\pm1)\}\subset\R^4,
\end{equation}
which one can think of as two copies of the standard $4$-QAM constellation.  Rotations of this constellation have been studied in DVB-NGH \cite{DVBNGH}, the outcome of which is summarized in \cite{douillard}, but only rotations of the form
\begin{equation}\label{dvbrot}
Q = \left(\begin{array}{rrrr}
a & -b & -b & -b \\
b & a & -b & b \\
b & b & a & -b \\
b & -b & b & a
\end{array}\right)
\end{equation}
were considered.  The orthogonality constraint on the rotation matrix forces $a^2 + 3b^2 = 1$, reducing the above matrix to being determined by only one parameter, the \emph{rotation parameter} $r := 3b^2/a^2$.  One can then solve for $a$ and $b$ given a fixed value of $r\in[0,1]$.  The DVB-NGH standard chose $r=0.4$ in order to minimize the bit error rate at the demapper \cite{douillard}.  Figure 2 is a plot comparing the codeword error rate  of the DVB-recommended rotation with the rotations determined by our method.

\begin{figure}[h!]
\hspace{-.02\textwidth}
\includegraphics[width = .5\textwidth]{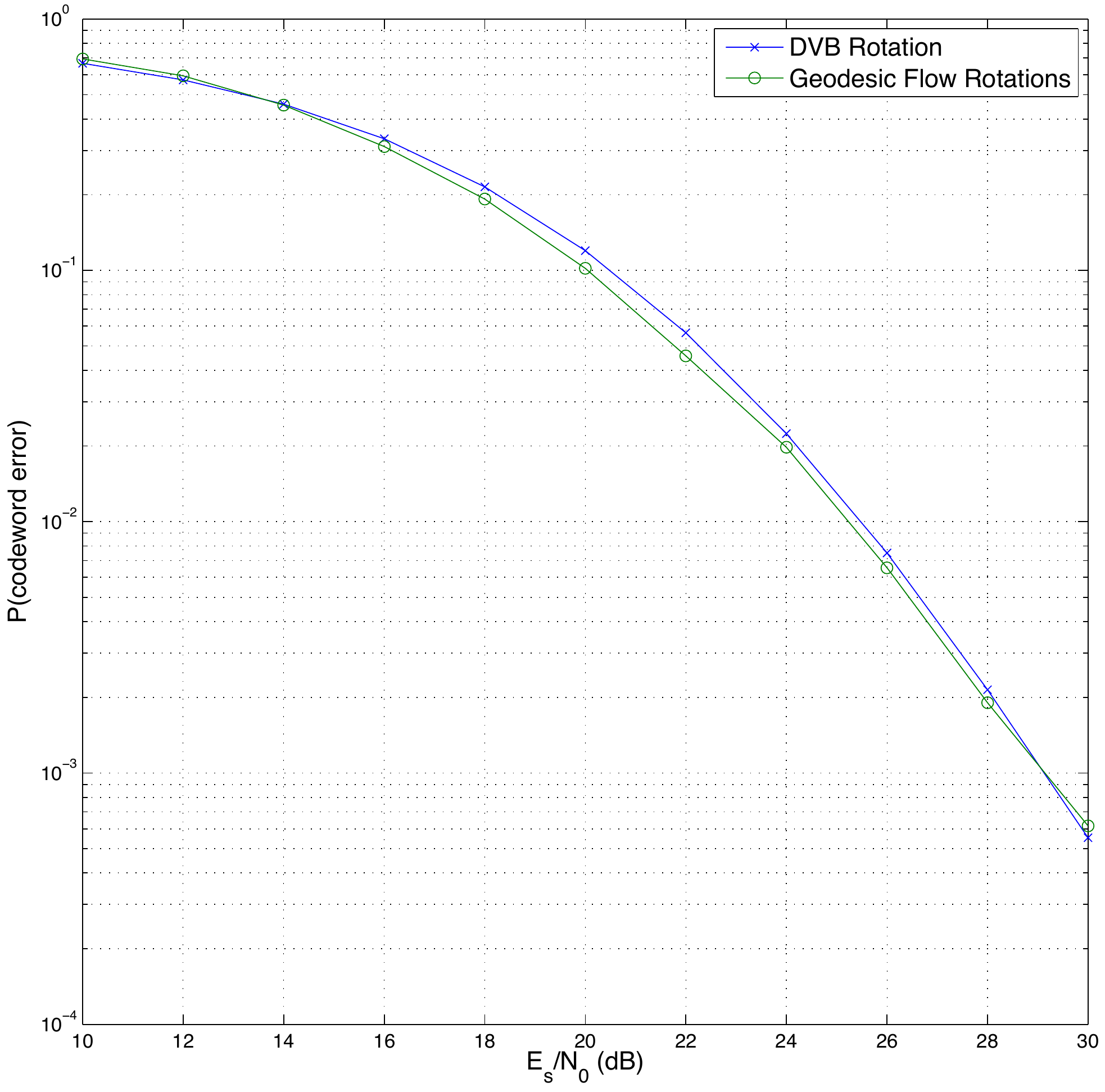}
\vspace{-.04\textwidth}\caption{Codeword Error Rates for $4$D $4$-QAM Constellations}
\end{figure}

Between $12$dB and $28$dB one can observe a slight improvement over the DVB-recommended rotation.  While the DVB rotation is only optimized over rotations of the form (\ref{dvbrot}), our algorithm produces, for example, the rotation
\begin{equation*}\label{ourrot}
Q_{24\text{dB}} = \left(\begin{array}{rrrr}
0.6253  &  0.1854  &  0.3542  &  0.6702 \\
-0.6702 &   0.6253 &   0.1854 &   0.3542 \\
-0.3542 &  -0.6702 &   0.6253 &   0.1854 \\
-0.1854 &  -0.3542 &  -0.6702 &   0.6253
\end{array}\right)
\end{equation*}
at an SNR of $24$dB.  This rotation matrix, as well as the rotations generated between $24$dB and $30$dB, takes the form
\begin{equation}\label{ourgenrot}
\left(\begin{array}{rrrr}
a & b & c & d \\
-d & a & b & c \\
-c & -d & a & b \\
-b & -c & -d & a
\end{array}\right)
\end{equation}
on which the orthogonality constraint forces the conditions
\begin{eqnarray*}
a^2 + b^2 + c^2 + d^2 &=& 1\\
ab - ad + cd + bc &=& 0.
\end{eqnarray*}
Matrices of the form (\ref{ourgenrot}) form a $2$-dimensional submanifold of $SO(4)$ over which our algorithm finds a local minimum.  This simulation demonstrates the benefit of optimizing over this $2$-dimensional submanifold of $SO(4)$, rather than the $1$-dimensional submanifold of matrices of the form (\ref{dvbrot}) considered by DVB.

\subsection{Algebraic Constellations}

Let us recall briefly how one constructs fully-diverse constellations from algebraic lattices, as described in \cite{OV}.  Starting with a totally real number field $K/\Q$ of degree $n$, one embeds the ring of integers $\mathcal{O}_K\hookrightarrow \R^n$ as a lattice, then searches for a principal ideal $I=(\alpha)\subseteq \mathcal{O}_K$ whose image under the above embedding is a rotated version of the orthogonal lattice $\Z^n$.  Full diversity of the corresponding ideal lattice is guaranteed by the fact that the number field is totally real.

One commonly used family of number fields are the cyclotomic fields $\Q(e^{2\pi i/m})$, for whose totally real subfields $\Q(\theta)$, $\theta = 2\cos(2\pi/m)$ there exists a standard way to find the aforementioned principal ideal.  One can see \cite{OV} for details on the construction, and many examples are catalogued at \cite{viterbowebsite}.  We examine how our method performs against corresponding cyclotomic constructions for $m=11$ and $m=17$, which result in lattices in $\R^5$ and $\R^8$, respectively.  

\begin{exam}
Consider the cyclotomic field $\Q(e^{2\pi i/11})$ and its totally real subfield $K=\Q(\theta),$ where $\theta = 2\cos(2\pi/11)$.  As $[K:\Q]=5$, the ring of integers of $K$ forms a full-rank lattice in $\R^5$.  The ideal lattice generated by the element $\alpha = 2-\theta$ has generator matrix $M = $
\begin{equation*}
\left(\begin{array}{rrrrr}
-0.1698 & -0.3260 & -0.4557 & -0.5485 & -0.5968 \\
-0.4557 & -0.5968 & -0.3260 & 0.1698 & 0.5485 \\
-0.5968 & -0.1698 & 0.5485 & 0.3260 & -0.4557 \\
-0.5485 & 0.4557 & 0.1698 & -0.5968 & 0.3260 \\
-0.3260 & 0.5485 & -0.5968 & 0.4557 & -0.1698 
\end{array}\right)
\end{equation*}
which can be found at \cite{viterbowebsite}. 

We compare two methods of rotating the constellation $\mathcal{C}=\{(\pm1,\pm1,\pm1,\pm1,\pm1) \}\subset \R^5$, namely using the above rotation matrix $M$, 
and using our numerical method.  That is, we compare the performance of
\begin{equation}
\mathcal{C}_{\text{cyclo}} = \{Mu : u\in \mathcal{C}\}
\end{equation}
with the rotations obtained using the geodesic flow algorithm between $20$dB and $28$dB.  While a plot of the respective PEP curves reveals that the resulting codes perform nearly identically, the data shows that our method slightly outperforms the algebraic construction.  One can see this in the below table:
\begin{comment}
\begin{figure}[h!]
\hspace{-.02\textwidth}
\includegraphics[width = .5\textwidth]{5D4QAMvsCYCLO.pdf}
\vspace{-.04\textwidth}\caption{Codeword Error Rates for $5$D Constellations}
\end{figure}
\end{comment}
\begin{equation*}
\begin{array}{|c|c|c|}
\hline
E_s/N_0&  \text{$P$(codeword error)}, & \text{$P$(codeword error)}, \\
\text{(dB)} & \mathcal{C}_{\text{cyclo}} & \text{GF Rotations}\\
\hline
20 & 0.2205 & 0.2183 \\
\hline
21 & 0.1610 & 0.1594 \\
\hline
22 & 0.1120 & 0.1108 \\
\hline
23 & 0.0739 & 0.0728 \\
\hline
24 & 0.0460 & 0.0454 \\
\hline
25 & 0.0273 & 0.0269 \\
\hline
26 & 0.0153 & 0.0151 \\
\hline
27 & 0.0082 & 0.0081 \\
\hline
28 & 0.0041 & 0.0041 \\
\hline
\end{array}
\end{equation*}

Our method essentially reproduces the generating matrix of the cyclotomic field, up to a numerical perturbation which accounts for the current SNR and the finiteness of the constellation.  For example, the geodesic flow algorithm produces the matrix $Q_{30\text{dB}} = $
\begin{equation*}
 \left(
\begin{array}{rrrrr}
    0.5842 &  -0.1856 &   0.3296 &   0.5550 &   0.4556 \\
   -0.4556 &   0.5842 &  -0.1856 &   0.3296 &   0.5550 \\
   -0.5550  & -0.4556 &   0.5842 &  -0.1856 &   0.3296 \\
   -0.3296   &-0.5550 &  -0.4556 &   0.5842 &  -0.1856 \\
    0.1856 &  -0.3296 &  -0.5550 &  -0.4556 &   0.5842
\end{array}\right)
\end{equation*}
at an SNR of $30$dB.  After multiplying appropriate rows and columns by $\pm1$ and swapping rows and columns (processes which do not change the underlying constellation), the matrix $Q_{30\text{dB}}$ can be put in a form which resembles the above matrix $M$ quite closely (in the sense, for example, that the Frobenius norm of their difference is small).

\end{exam}

\begin{exam}
To demonstrate the ability of our algorithm to perform numerical analysis on relatively high-dimensional constellations, let us perform a similar experiment in $\R^8$.  We note that parametrizing rotation matrices in $\R^8$ directly would require defining and numerically optimizing over $\dim SO(8) = (8^2-8)/2 = 28$ variables.

We consider the constellation
\begin{equation*}
\mathcal{C}=\{(\pm1,\ldots,\pm1) \}\subset \R^8
\end{equation*}
which we rotate according to generator matrix of the ideal lattice coming from the totally real subfield of the cyclotomic field $\Q(e^{2\pi i/17})$.  The totally real subfield is then $K=\Q(\theta)$, where $\theta=2\cos(2\pi/17)$, and the corresponding principal ideal is generated by the element $2-\theta$.

Instead of numerically computing a different rotation at each SNR, we will use only the rotation matrix $Q_{30\text{dB}}$
\begin{comment}
\footnote{The matrix generated by the geodesic flow algorithm, for the constellation $\mathcal{C}=\{(\pm1,\ldots,\pm1) \}\subset \R^8$ at an SNR of $30$dB is
\begin{equation*}
Q_{30\text{dB}} = 
\left(\begin{array}{rrrrrrrr}
0.3289 & -0.4247 & -0.3690 & 0.3288 & 0.4307 & 0.3415 & 0.1324 & 0.3839 \\
-0.3901 & 0.4210 & -0.4348 & -0.3459 & 0.3175 &  0.3688 &  0.3325 & -0.1206 \\
-0.1324 & -0.3839 &   0.3289 & -0.4247 &  -0.3690 & 0.3288 & 0.4307 & 0.3415 \\
-0.3325 &   0.1206 &  -0.3901 &   0.4210 &  -0.4348 &  -0.3459 &   0.3175 &   0.3688 \\
-0.4307 &  -0.3415 &  -0.1324 &  -0.3839 &  0.3289 &  -0.4247 &  -0.3690 &   0.3288 \\
-0.3175 &  -0.3688 &  -0.3325 &  0.1206 &  -0.3901 &   0.4210 &  -0.4348 &  -0.3459 \\
0.3690 &  -0.3289 &  -0.4307 &  -0.3415 &  -0.1324 &  -0.3839 &   0.3289 &  -0.4247 \\
0.4348 &   0.3459 &  -0.3175 &  -0.3688 & -0.3325 &   0.1206 &  -0.3901 &  0.4210
\end{array}\right)
\end{equation*}
\twocolumn}
\end{comment}
computed by our algorithm, over a larger range of SNR values.  This is perhaps a more practical scenario; it requires the use of only one codebook, rather than a different one for each SNR.  For this simulation, we also change the step size to $h=.001$.  A plot of the relative performance of the constellations is presented in Figure 3.
\begin{figure}[h!]
\hspace{-.02\textwidth}
\includegraphics[width = .5\textwidth]{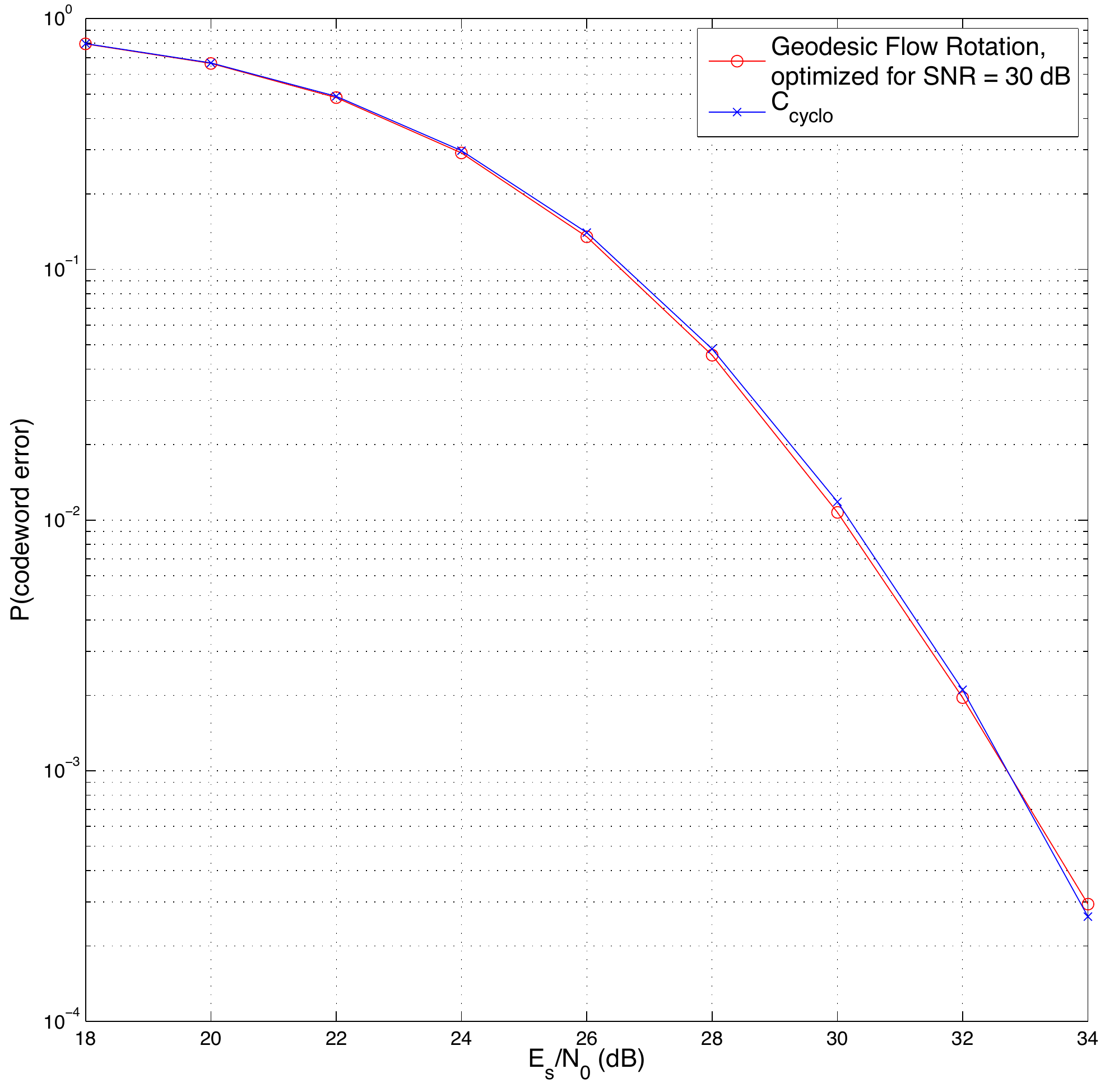}
\vspace{-.04\textwidth}\caption{Codeword Error Rates for $8$D Constellations}
\end{figure}

While the numerical constellation is optimized only for $30$dB, close examination of the data reveals that it outperforms the cyclotomic constellation between $18$dB and $32$dB.  Furthermore, the largest gain in performance occurs at a codeword error rate of approximately $10^{-2}$, when the SNR is approximately $30$dB, as expected.

\end{exam}

\section{Conclusions and Future Work}\label{conclusions}

In this paper, we presented a numerical algorithm for rotating constellations based on a geodesic flow algorithm on the Lie group $SO(n)$ of all $n$-dimensional rotation matrices.  The algorithm converges to a local minimum of an upper bound on the pairwise error probability.  Our examples show explicitly how our method improves the performance of the constellation, both against non-rotated constellations, and constellations which have been rotated by previously known methods.

While we have focused on only a single objective function in our paper, there are others relevant for codebook design for Rayleigh fading channels.  For instance, mutual information is often used as a design criteria, and we hope to apply our method to numerically optimize the mutual information of a constellation over all possible rotations.

\section{Acknowledgements}

The first author has been partially supported by a grant from the Magnus Ehrnrooth Foundation.

\bibliographystyle{ieee}
\bibliography{myrefs_new}

\end{document}